\documentclass{elsart}

\usepackage{bm}
\usepackage{amsmath}
\usepackage{graphicx}

\newcommand{\mnp}{\mathnormal{\Psi}}
\newcommand{\mng}{\mathnormal{\Gamma}}
\newcommand{\mno}{\mathnormal{\Omega}}
\newcommand{\sgn}{\mathrm{sgn}}

\begin{document}

\begin{frontmatter}

\title{Modulational instabilities in neutrino--anti-neutrino
interactions}   

\author[Goteborg]{Mattias Marklund},
\author[Bochum,Umea]{Padma K.\ Shukla},
\author[Goteborg,CapeTown]{Gerold Betschart},
\author[Umea]{Lennart Stenflo},
\author[Goteborg]{Dan Anderson}
and
\author[Goteborg]{Mietek Lisak}

\address[Goteborg]{Department of Electromagnetics, Chalmers University of
Technology, \\SE--412 96 G\"oteborg, Sweden}
\address[Bochum]{Institut f\"ur Theoretische Physik IV, Fakult\"at f\"ur
  Physik und Astronomie, Ruhr-Universit\"at Bochum, D--44780 Bochum,
  Germany} 
\address[Umea]{Department of Plasma Physics, Ume{\aa} University,
  SE--901 87 Ume{\aa}, Sweden}
\address[CapeTown]{Department of Mathematics and Applied Mathematics,
  University of Cape Town, 7701 Rondebosch, Cape Town, South Africa}

\date{\today}

\begin{abstract}
We analyze the collective behavior of neutrinos and antineutrinos in a 
dense background. Using the Wigner transform technique, it is shown that the 
interaction can be modelled by a coupled system of nonlinear Vlasov-like 
equations. From these equations, we derive a dispersion relation for 
neutrino-antineutrino interactions on a general background. The
dispersion relation admits a novel modulational instability. 
The results are examined, together with a numerical example, and we discuss the
induced density inhomogeneities using parameters relevant to the early
Universe.  

\end{abstract}

\end{frontmatter}

\section{Introduction}

Neutrinos have fascinated people ever since they were first introduced
by Wolfgang Pauli in 1931. Since then, neutrinos have gone from
hypothetical to an extremely promising tool for analysing
astrophysical events, and neutrino cosmology is now one of the hottest
topics in modern time due to the discovery that neutrinos
may be massive \cite{ahmad}. Because of its weak interaction with other
particles, neutrinos can travel great distances without being affected
appreciably by material obstacles. They can therefore give us detailed 
information about events taking place deep within, e.g.  supernov{\ae}. 
Furthermore, since the neutrinos decoupled from matter at a redshift
$z$ of the order $10^{10}$, as compared to $z \simeq 10^3$ for photons, 
it is possible that neutrinos could give us a detailed understanding
of the early Universe, if such a signal could be detected \cite{Dolgov}. 
Massive neutrinos have also been a possible candidate for hot dark matter 
necessary for explaining certain cosmological observations, such as rotation
curves of spiral galaxies \cite{Sofue-Rubin}. Thus, massive neutrinos
could have a profound influence on the evolution of our Universe. 
Unfortunately, due to the Tremaine--Gunn bound \cite{Tremaine-Gunn}, 
the necessary mass of the missing particles (if fermions) for explaining 
the formation of dwarf galaxies seems to make neutrinos of \emph{any} 
species unlikely single candidates for dark matter. As a remedy to this 
problem, interacting hot dark matter has been suggested
\cite{Raffelt-Silk,Atrio-Barandela-Davidson}, since the interaction
prevents the free-streaming smoothing of small scale neutrino
inhomogeneities. Thus, dark matter in astrophysics is not only 
a mystery but it also plays an essential role in determining  
the dynamics of the universe, its large scale structures,
the galaxies and superclusters. However, so far, the suggested 
``sticky'' neutrino models have not been successful in dealing 
with the dwarf galaxy problem \cite{Raffelt-Silk}.   

A first successful indication that neutrinos have a non-zero mass
came in 1998 through laboratory experiments of atmospheric neutrinos
and their oscillations \cite{Fukuda}. Although the allowed neutrino
masses encompass a wide range\footnote{Some estimates even support
the notion that neutrinos may contribute up to 20\% of the matter 
density of the Universe \cite{Elgaroy}.}, it is currently believed that 
neutrinos have masses below 2~eV. This conclusion 
is furthermore supported by independent cosmological observations 
(see, e.g., \cite{Hannestad}). Thus, the masses of neutrinos are indeed 
very small, and the classical analysis by Tremaine and Gunn would 
thereby indicate that neutrinos can in no way be considered as a sole 
candidate for dark matter. This conclusion will in this paper be re-analyzed 
within the electro-weak framework, where neutrino--neutrino interactions 
occur as a natural consequence of the theory.

Thus, in this Letter, we consider the nonlinear interaction between neutrinos 
and antineutrinos in the lepton plasma of the early Universe, adopting a 
semi-classical model. Neutrinos and antineutrinos interact with dense
plasmas through the charged and neutral weak currents arising from
the Fermi weak nuclear interaction forces. Charged weak currents involve the 
exchange of the charged vector bosons associated with the processes
involving interactions between leptons and neutrinos of the
same flavor, while neutrino weak currents involve the exchange 
of the neutral vector bosons associated with processes involving 
neutrinos of all types interacting with arbitrary charged and
neutral particles. Asymmetric flows of neutrinos and antineutrinos 
in the early Universe plasma may be created by the ponderomotive
force  of nonuniform intense photon beams or by shock waves. 
Here, using an effective field theory approach, a system of coupled 
Wigner-Moyal equations for nonlinearly interacting neutrinos and 
antineutrinos is derived, and  it is shown that these 
equations admit a modulational instability.  Finally, we discuss the 
relevance of our results in the context of the dark matter problem, 
and it is moreover suggested that the nonlinearly excited
fluctuations could be used as a starting point for obtaining a better
understanding of the process of galaxy formation. It turns out that 
the short-time evolution of the primordial neutrino plasma medium in the 
temperature range $1 \,{\rm MeV} < T < 10 \,{\rm MeV}$ is governed by 
collisionless collective effects involving relativistic neutrinos
and antineutrinos.

\section{Dispersion relation and the motion of neutrino bunches}

As a primer, we will study the implication of the known dispersion of
neutrinos on a thermal neutrino/anti-neutrino background, using the
eikonal representation and the WKBJ approximation. 

Suppose that a single neutrino (or anti-neutrino) moves in a Fermionic 
sea composed of neutrino--antineutrino admixture. The energy $E$ of
the neutrino (antineutrino) is then given by (see, e.g. 
\cite{Kuo-Pantaleone,Silva-etal})
\begin{equation}\label{energy}
E = \sqrt{p^2c^2 + m^2c^4} + V_{\pm}(\bm{r},t) ,
\end{equation}
where $\bm{p}$ is the neutrino (anti-neutrino) momentum, $c$ the
speed of light in vacuum, and $m$ the neutrino mass. The effective
potential for a neutrino moving on a background of it's own flavor 
and in thermal equilibrium is given by\footnote{For a more detailed 
description of the potential, see the next section.}
\cite{Kuo-Pantaleone} (see also 
\cite{Weldon,Notzhold-Raffelt,Nunokawa-etal,Raffelt}) 
\begin{subequations}
\begin{equation}
V_{\pm}(\bm{r},t) = \pm 2\sqrt{2}G_F (n - \bar{n}) ,
\label{potential1}
\end{equation}
while the potential for a neutrino moving on a background of a different
flavor is
\begin{equation}
V_{\pm}(\bm{r},t) = \pm \sqrt{2}G_F (n - \bar{n}) ,
\label{potential2}
\end{equation}
\label{potential}
\end{subequations}
where $G_F/(\hbar c)^3  \approx 1.2\times 10^{-5}\,
\mathrm{GeV}^{-2}$, $G_F$ is the Fermi constant, $n$ ($\bar{n}$) 
is the density of the background neutrinos (antineutrinos), and $+$
($-$) represents the propagating neutrino (antineutrino). Expressions
(\ref{potential}) are valid in the rest frame of the background.
As seen from (\ref{energy}) and (\ref{potential}), while neutrinos
moving in a background of neutrinos and antineutrinos change their energy 
by an amount $\sim G_F(n - \bar{n})$, the antineutrinos change
their energy by $\sim-G_F(n - \bar{n})$ \cite{Tsintsadze-etal}. The
extra factor of $2$ in (\ref{potential1}) as compared to expression
(\ref{potential2}) comes from exchange effects between identical
particles \cite{Notzhold-Raffelt}. 
  
The relation (\ref{energy}) can be interpreted as a dispersion
relation for relativistic and nonrelativistic neutrinos, with 
the identifications $E = \hbar\omega$ and $\bm{p} = \hbar\bm{k}$, i. e.
\begin{equation}\label{disprel}
\omega = c\sqrt{k^2 + \frac{m^2c^2}{\hbar^2}} + \frac{V_{\pm}}{\hbar} ,
\end{equation}
where $\hbar$ is the Planck constant divided by $2\pi$. By using 
the eikonal representation (viz.\ $E \rightarrow \hbar \omega_0 -i\hbar
\partial/\partial t$ and $\bm{p}\rightarrow \hbar \bm{k}_0 + i 
\hbar \nabla$) and the WKBJ approximation \cite{Karpman,Hasegawa} 
(viz.\ $|\partial{\mnp}/\partial t| \ll \omega_0|{\mnp}|$ and 
$|\nabla{\mnp}| \ll |\bm{k}_0||{\mnp}|$), we obtain from 
Eq.~(\ref{disprel}) a Schr\"odinger equation for slowly varying modulated 
(by long-scale density fluctuations) neutrino (anti-neutrino) 
wave function ${\mnp}(\bm{r}, t)$ (i.e., neutrino bunches) 
(see also Ref.\ \cite{Tsintsadze-etal}  
for a similar treatment of neutrino--electron interactions)
\begin{equation} \label{maineq}
  i\left( \frac{\partial}{\partial t} +
  \bm{v}_g\cdot\bm{\nabla}\right){\mnp} 
  + \frac{\hbar}{2\gamma m}\left[
  \nabla^2 - \frac{(\gamma^2 -
  1)}{\gamma^2}(\bm{n}_0\cdot\bm{\nabla})^2 
  \right]{\mnp} -
  \frac{V_{\pm}}{\hbar}{\mnp} = 0 ,
\end{equation}
where $\bm{v}_g = c\bm{k}_0(k_0^2 + m^2c^2/\hbar^2)^{-1/2}$ is the
group velocity \footnote{We note that when the scalelength
of the density inhomogeneity is comparable to the wavelength of the
modulated neutrino wave packets, we must modify the coupled
Schr\"odinger equations to account for differing group velocities of
neutrinos and antineutrinos in a Fermionic sea. We expect a shift in
the momentum of Eq.\ (\ref{Wigner}) and a slower growth rate of the
modulational instability of neutrino quasi-particles involving short-scale
density inhomogeneities.}  
of relativistic neutrinos and antineutrinos which have similar energy 
spectra, $\gamma = (1 - v_g^2/c^2)^{-1/2}$ is the relativistic 
gamma factor, $\bm{n}_0 = \bm{k}_0/|k_0|$, and $\bm{k}_0$ is the
vacuum wavevector. Suppose now that the neutrino bunches themselves
are nearly in thermal equilibrium (to be quantified in the next
section). Then, we have the case of self-interacting neutrinos and
anti-neutrinos, and the densities in the potential $V_{\pm}$ is given
in  terms of the sums
\begin{equation}
  n = \sum_{i = 1}^{M} n_i = \sum_{i =
  1}^{M}\langle|{\mnp}_{i+}|^2\rangle , \,  
  \bar{n} = \sum_{i = 1}^N \bar{n}_i = 
  \sum_{i = 1}^N \langle|{\mnp}_{i-}|^2\rangle , 
\label{densities}
\end{equation}
where ${\mnp}_{i+}$ and ${\mnp}_{i-}$ are the neutrino and
antineutrino wave functions (with $i$ numbering the wave functions),
respectively, and the angular bracket denotes the ensemble average.  
In this case, the relativistic neutrino and antineutrino wave packets are 
comoving with the background, and Eq.\ (\ref{maineq}) thus yields 
\begin{equation}
  i\frac{\partial{\mnp}_{i\pm}}{\partial t} +
  \frac{\hbar}{2m\gamma}\left( \nabla_{\perp}^2 +
  \frac{1}{\gamma^2}\nabla_{||}^2 \right){\mnp}_{i\pm} -
  \frac{V_{\pm}}{\hbar}{\mnp}_{i\pm} = 0 ,
\label{modnlse}
\end{equation} 
where $
  \nabla_{\perp}^2 = \nabla^2 - (\bm{n}_0\cdot\bm{\nabla})^2 , \quad 
  \nabla_{||}^2    = (\bm{n}_0\cdot\bm{\nabla})^2$. 
Expressions (\ref{potential1}) and (\ref{densities}) reveal that
self-interactions between relativistic neutrinos and antineutrinos produce 
a nonlinear asymmetric potential in Eq.\ (\ref{modnlse}). By further rescaling 
the coordinate along $\bm{n}_0$, Eq.\ (\ref{maineq}) can finally be written as 
the coupled system 
\begin{equation}
  i\frac{\partial{\mnp}_{i\pm}}{\partial t} +
  \frac{\alpha}{2}\nabla^2{\mnp}_{i\pm} \mp
  \beta(n - \bar{n}){\mnp}_{i\pm} = 0 ,  
\label{interaction}
\end{equation}
where $\alpha = \hbar/m\gamma$, and $\beta =
2\sqrt{2}G_F/\hbar$ for neutrinos moving on the
same flavor background.  

Equation (\ref{interaction}) shows that this approach can
lead to some interesting effects. The case of a single
self-interacting neutrino bunch shows that the formation of dark
solitary structures is possible. Furthermore, the slightly more
complicated case of two interacting bunches, either of the
neutrino--neutrino or neutrino--anti-neutrino type, can result in
splitting and focusing of the wave packets \cite{Marklund-Shukla-Stenflo}.

\section{Kinetic description}

In the preceding section, we investigated the case of a neutrino
bunch close to thermal equilibrium. In general, this may of
course not be the case, and Eq.\ (\ref{potential}) must be modified. 
The more precise form of the potential  $V_{\pm}$ for equal species 
due to neutrino forward scattering is given by 
\cite{Pantaleone} 
\begin{equation} \label{fullpotential}
  V_{\pm}(t,\bm{r},\bm{p};f_{i\pm}) =  \pm2\sqrt{2}\,G_F\int d\bm{q}\, 
  (1 - \hat{\bm{p}}\cdot\hat{\bm{q}})\left[\sum_{i =1}^M
  f_{i+}(t,\bm{r},\bm{q}) - \sum_{i = 1}^N f_{i-}(t,\bm{r},\bm{q})
  \right] ,   
\end{equation}
where hatted quantities denote the corresponding unit vector, and
$f_{i+}(t,\bm{r},\bm{q})$ ($f_{i-}(t,\bm{r},\bm{q})$) is the neutrino
(anti-neutrino) distribution function corresponding to bunch $i$. The
distribution functions are defined to be normalized such that 
\begin{equation}
  n_i(t,\bm{r}) = \int\,d\bm{q}\,f_{i+}(t,\bm{r},\bm{q}), \quad 
  \bar{n}_i(t,\bm{r}) = \int\,d\bm{q}\,f_{i-}(t,\bm{r},\bm{q}),
\end{equation}
where $n_i$ ($\bar{n}_i$) is the number density of the $i^\text{th}$
neutrino (anti-neutrino) bunch.

The first thing to notice is that when the distribution is thermal,
the potential (\ref{fullpotential}) reduces exactly to (\ref{potential1}). 
Secondly, when the neutrinos have an almost thermal distribution, i.e.  
the corresponding distribution function may be expressed as (dropping
the indices for notational simplicity) $f(t,\bm{r},\bm{p}) = f_0(p) + \delta\!
f(t,\bm{r},\bm{p})$, where $|\delta\!f| \ll |f_0|$, we obtain the 
following form of the potential 
\begin{equation}
  V_{\pm}(t,\bm{r},\bm{p};f_{i\pm}) = \pm2\sqrt{2}\,G_F\left[ (n - \bar{n}) -
  \int\,d\bm{q}\,(\hat{\bm{p}}\cdot\hat{\bm{q}})\left(\sum_{i
  = 1}^M\delta\!f_{i+} - \sum_{i = 1}^N \delta\!f_{i-}\right) \right] .  
\end{equation}
The last term is small and may therefore be neglected, and we obtain
$V_{\pm}(t,\bm{r}) \approx \pm2\sqrt{2}\,G_F (n - \bar{n})$, 
in accordance with expressions (\ref{potential1}), thus justifying the
equation of motion (\ref{interaction}).

Now, we define a distribution function for the neutrino states 
by Fourier transforming the two-point correlation function for
${\mnp}_{\pm}$, according to \cite{Wigner} 
\begin{equation}
  {f}_{i\pm}( t, \bm{r}, \bm{p}) =
  \frac{1}{(2\pi\hbar)^3}\int\,d\bm{y}\,
  \mathrm{e}^{i\bm{p}\cdot\bm{y}/\hbar}   
  \langle {\mnp}_{i\pm}^*( t, \bm{r} +
  \bm{y}/2){\mnp}_{i\pm}( t,  \bm{r} - \bm{y}/2)\rangle ,
\label{statdef}
\end{equation}
where $\bm{p}$ represents the momentum of the neutrino (antineutrino)
quasi-particles (note that the ensemble average was not present in the
original definition \cite{Wigner}, but has important consequences when
the phase of the wave function has a random component \cite{Hall-etal}; 
for similar treatments of optical beams and quantum
plasmas, see \cite{Hall-etal} and \cite{Anderson-etal}, respectively). 
We note that with the definition (\ref{statdef}), the following relation
holds 
\begin{equation}
  \langle |{\mnp}_{i\pm}( t, \bm{r})|^2\rangle =
  \int\,d\bm{p}\,{f}_{i\pm}( t, \bm{r}, \bm{p}) .
\end{equation}
Thus, using (\ref{statdef}) and (\ref{modnlse}) together with the
potential (\ref{fullpotential}) we obtain the generalized
Wigner--Moyal equation 
\begin{equation}
  \frac{\partial{f}_{i\pm}}{\partial t}
  + \frac{\bm{p}}{m\gamma}\cdot\frac{\partial{f}_{i\pm}}{\partial\bm{r}} -
  \frac{2V_{\pm}}{\hbar}\sin\left[
  \frac{\hbar}{2}\left(\overleftarrow{\frac{\partial}{\partial\bm{r}}}
  \cdot 
  \overrightarrow{\frac{\partial}{\partial\bm{p}}} 
  - \overleftarrow{\frac{\partial}{\partial\bm{p}}} \cdot 
  \overrightarrow{\frac{\partial}{\partial\bm{r}}} \right)
  \right]{f}_{i\pm}  = 0 ,  
\label{Wigner}
\end{equation}
for $f_{i\pm}$, where the $\sin$-operator is defined in
terms of its Taylor expansion, and the arrows denote the direction of
operation. In the case of the potential (\ref{potential1}), the last
term in the $\sin$-operator drops out, and Eq.\
(\ref{Wigner}) reduces to the standard Wigner--Moyal equation
\cite{Wigner}. 

Retaining only the lowest order terms in $\hbar$ (i.e. taking the
\emph{long wavelength} limit), we obtain the coupled Vlasov
equations  
\begin{equation}
  \left[\frac{\partial}{\partial t}
  +\left(\frac{\bm{p}}{m\gamma} +
  \frac{\partial V_{\pm}}{\partial\bm{p}}
  \right)\cdot\frac{\partial}{\partial\bm{r}}\right]{f}_{i\pm}  
  - \frac{\partial V_{\pm}}{\partial\bm{r}}\cdot
  \frac{\partial{f}_{i\pm}}{\partial\bm{p}} = 0 .
\label{Vlasov}
\end{equation}
The term $\partial V_{\pm}/\partial\bm{p}$ represents the group
velocity. 
While higher order group velocity dispersion is
present in (\ref{Wigner}), this is not the case in
(\ref{Vlasov}). Thus, information is partially lost by using the Eq.\
(\ref{Vlasov}). Furthermore,   
while Eq.~(\ref{Vlasov}) preserves the number of quasi-particles,
Eq.~(\ref{Wigner}) shows that this conclusion is in general not true,
i.e. the particle number in a phase-space volume is not
constant, and the higher order terms $\partial^n
V_{\pm}/\partial\bm{r}^n$ may moreover contain vital short wavelength
information. Equations similar to (\ref{Vlasov}) have been used to study
neutrino--electron interactions in astrophysical contexts
\cite{Silva-etal}. 

Suppose now that we have small amplitude perturbations on a background of
constant neutrino and antineutrino densities $n_i = n_{i0}$ and
$\bar{n}_i = \bar{n}_{i0}$, respectively, i.e.
\begin{equation}
  {f}_{i\pm}( t, \bm{r}, \bm{p}) = {f}_{i0\pm}(\bm{p}) +
  \delta\!{f}_{i\pm}(\bm{p})\exp[i(\bm{K}\cdot\bm{r} -
  {\mno} t)] , 
\end{equation}
and $|\delta\!{f}_{i\pm}| \ll |{f}_{i0\pm}|$, where $\bm{K}$
and ${\mno}$ is the perturbation wavevector and frequency,
respectively. Thus, Eqs.\ (\ref{Wigner}) give
\begin{eqnarray}
  && i\left[{\mno} -\frac{\bm{p}\cdot\bm{K}}{m\gamma} - 
  \frac{2i}{\hbar}V_{0\pm}\sin\left(-\frac{i\hbar}{2}%
    \overleftarrow{\frac{\partial}{\partial\bm{p}}}%
    \cdot\bm{K}\right)
  \right]\delta\!{f}_{i\pm} \nonumber \\
  && \qquad\qquad\qquad\qquad+
  \frac{2}{\hbar}\delta\!V_{\pm}\sin\left(\frac{i\hbar}{2}\bm{K}%
    \cdot\overrightarrow{\frac{\partial}{\partial\bm{p}}}%
  \right){f}_{i0\pm} = 0 , 
\label{cw}
\end{eqnarray}
where $\delta\!V_{\pm} = V_{\pm}(t,r,\bm{p};\delta\!f_{i\pm})$ and
$V_{0\pm} = V_{\pm}(t,r,\bm{p};f_{i0\pm})$ from
Eq.\ (\ref{fullpotential}). 
Eliminating $\delta\!f_{i\pm}$ 
from (\ref{cw}), using $\delta\!V_{-} = -\delta\!V_{+}$, we have
\begin{eqnarray}
  \delta\!V_{+}(\bm{p}) &=& \frac{4\sqrt{2}\,i
  G_F}{\hbar}\int\,d\bm{q}\,(1 -
  \hat{\bm{p}}\cdot\hat{\bm{q}})\,\delta\!V_+(\bm{q}) \nonumber \\
  && \times \left[
  \sum_{i = 1}^M \frac{\sin\left(
  \frac{i\hbar}{2}\bm{K}\cdot\overrightarrow{\frac{\partial}{\partial\bm{q}}}
  \right)f_{i0+}(\bm{q})}{{\mno} -
  \frac{\bm{q}\cdot\bm{K}}{m\gamma} - 
  \frac{2i}{\hbar}V_{0+}(\bm{q})\sin\left(
  \frac{i\hbar}{2}\overleftarrow{\frac{\partial}{\partial\bm{q}}}\cdot\bm{K}
  \right)}  \right. \nonumber \\
  &&\qquad\quad \left.+ \sum_{i = 1}^N \frac{\sin\left(
  \frac{i\hbar}{2}\bm{K}\cdot\overrightarrow{\frac{\partial}{\partial\bm{q}}}
  \right)f_{i0-}(\bm{q})}{{\mno} -
  \frac{\bm{q}\cdot\bm{K}}{m\gamma} - 
  \frac{2i}{\hbar}V_{0-}(\bm{q})\sin\left(
  \frac{i\hbar}{2}\overleftarrow{\frac{\partial}{\partial\bm{q}}}\cdot\bm{K}
  \right)}
  \right]  .
\label{dispersionequation}
\end{eqnarray}
Assuming that $\delta\!f_{i\pm}(\bm{p})$ is a symmetric function of
$\bm{p}$ implies that $\delta\!V_{\pm}$ is independent of $\bm{p}$,
and Eq.\ (\ref{dispersionequation}) simplifies to the dispersion
relation 
\begin{eqnarray}
  1 &=& \frac{4\sqrt{2}\,i
  G_F}{\hbar}\int\,d\bm{q}\,\left[
  \sum_{i = 1}^M \frac{\sin\left(
  \frac{i\hbar}{2}\bm{K}\cdot{\frac{\partial}{\partial\bm{q}}}
  \right)f_{i0+}(\bm{q})}{{\mno} -
  \frac{\bm{q}\cdot\bm{K}}{m\gamma} - 
  \frac{2i}{\hbar}\sin\left(
  \frac{i\hbar}{2}\bm{K}\cdot{\frac{\partial}{\partial\bm{q}}}
  \right)V_{0+}(\bm{q})}  \right. \nonumber \\
  &&\qquad\quad \left.+ \sum_{i = 1}^N \frac{\sin\left(
  \frac{i\hbar}{2}\bm{K}\cdot{\frac{\partial}{\partial\bm{q}}}
  \right)f_{i0-}(\bm{q})}{{\mno} -
  \frac{\bm{q}\cdot\bm{K}}{m\gamma} - 
  \frac{2i}{\hbar}\sin\left(
  \frac{i\hbar}{2}\bm{K}\cdot{\frac{\partial}{\partial\bm{q}}}
  \right)V_{0-}(\bm{q})}
  \right]  ,
\label{dispersionrelation}
\end{eqnarray}
where we have dropped the arrows indicating the direction of
operation. Note that if the background distribution is thermal,
$V_{0\pm}$ is independent of $\bm{p}$, and the last term in the
denominators of Eq.\ (\ref{dispersionrelation}) vanishes.

\subsection{The one-dimensional case}

The simplest way to analyse the dispersion relation
(\ref{dispersionrelation}) is to reduce the dimensionality of the
problem. We therefore first look at the one-dimensional case, where we
may use the identity $
  2\sin\left(\frac{i\hbar K}{2} \frac{\partial}{\partial p}
  \right)h(p) = i\left[ h(p + \frac{\hbar K}{2}) -
  h(p - \frac{\hbar K}{2})\right]
$, in order to rewrite the dispersion relation (\ref{dispersionrelation}) 
as 
\begin{eqnarray}
  1 &=& -\frac{2\sqrt{2}\,G_F}{\hbar}\int dq \left\{
  \sum_{i = 1}^M \frac{f_{i0+}(q + \hbar K/2) - f_{i0+}(q - \hbar
  K/2)}{{\mno} - qK/m\gamma + \Delta_{+}(q)
  }  \right. \nonumber \\
  && \left.+ \sum_{i = 1}^N \frac{f_{i0-}(q + \hbar K/2) -
  f_{i0-}(q - \hbar K/2)}{{\mno} -
  qK/m\gamma + \Delta_{-}(q)} 
  \right\} ,
\label{dispersion2}
\end{eqnarray}
where we have introduced $\Delta_{\pm}(q) \equiv [V_{0\pm}(q + \hbar
  K/2) - V_{0\pm}(q - \hbar K/2]/\hbar$.
In the case of mono-energetic beams, i.e. ${f}_{i0+}(p) =
n_{i0}\delta(p - p_{i0})$ and ${f}_{i0-}(p) =
\bar{n}_{i0}\delta(p - \bar{p}_{i0})$, Eq.~(\ref{dispersion2}) reduces
to
\begin{eqnarray}
  1 &=& -\frac{2\sqrt{2}\, G_F}{\hbar}\Bigg\{ \sum_{i = 1}^M
  n_{i0}\left[-\frac{\hbar K^2}{m\gamma} + \Delta_+(p_{i0} + \hbar K/2) -
  \Delta_+(p_{i0} - \hbar K/2)\right] \nonumber \\
  &&  \times\bigg[\left({\mno} -
  \frac{p_{i0}K}{m\gamma}\right)^2 - \left(\frac{\hbar
  K^2}{2m\gamma}\right)^2 \nonumber \\
  &&\quad + \left({\mno} -
  \frac{p_{i0}K}{m\gamma}\right)[\Delta_+(p_{i0} + \hbar 
  K/2) + \Delta_+(p_{i0} - \hbar K/2)]  \nonumber \\
  &&\quad  + \frac{\hbar
  K^2}{2m\gamma}[\Delta_+(p_{i0} + \hbar K/2) - \Delta_+(p_{i0} -
  \hbar K/2)] \nonumber \\
  &&\qquad + \Delta_+(p_{i0} + \hbar K/2)\Delta_+(p_{i0} - \hbar
  K/2)\bigg]^{-1} \nonumber \\ 
  && + \sum_{i = 1}^N
  \bar{n}_{i0}\bigg[-\frac{\hbar K^2}{m\gamma} + \Delta_-(\bar{p}_{i0}
  + \hbar K/2) - 
  \Delta_-(\bar{p}_{i0} - \hbar K/2)\bigg]  \nonumber \\
  &&  \times\bigg[\left({\mno} -
  \frac{\bar{p}_{i0}K}{m\gamma}\right)^2 - \left(\frac{\hbar
  K^2}{2m\gamma}\right)^2 \nonumber \\
  &&\quad + \left({\mno} -
  \frac{\bar{p}_{i0}K}{m\gamma}\right)[\Delta_-(\bar{p}_{i0} + \hbar 
  K/2) + \Delta_-(\bar{p}_{i0} - \hbar K/2)]  \nonumber \\
  &&\quad  + \frac{\hbar
  K^2}{2m\gamma}[\Delta_-(\bar{p}_{i0} + \hbar K/2) -
  \Delta_-(\bar{p}_{i0} - 
  \hbar K/2)] \nonumber \\
  &&\qquad + \Delta_-(\bar{p}_{i0} + \hbar
  K/2)\Delta_-(\bar{p}_{i0} - \hbar 
  K/2)\bigg]^{-1} \Bigg\} .
\label{monodispersion}
\end{eqnarray}
where 
\begin{equation}
  V_{0\pm}(p) = \pm 2\sqrt{2}\,G_F\left[ (n_0 - \bar{n}_0) -
  {\sgn}(p)\left(\sum_{i = 1}^M n_{i0}\,{\sgn}(p_{i0}) - \sum_{i = 1}^N
  \bar{n}_{i0}\,{\sgn}(\bar{p}_{i0})\right) \right] ,
\label{monopotential}
\end{equation}
by Eq.\ (\ref{fullpotential}).

\begin{figure}
\begin{center}
  \includegraphics[width=.85\textwidth]{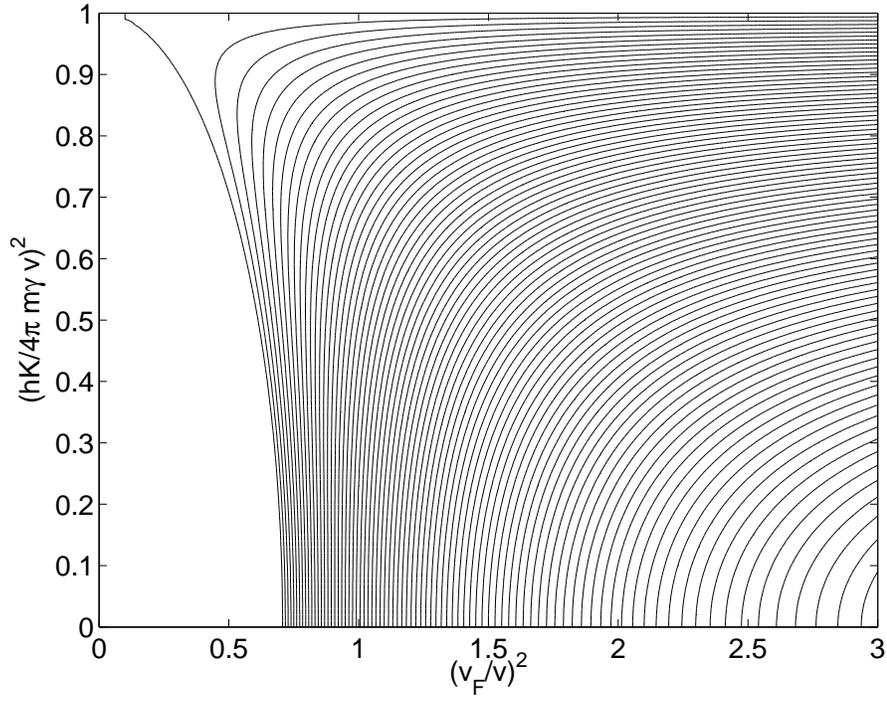}
\end{center}
  \caption{A contour plot of the values of
  $({\mng}/Kv)^2$, when $\varepsilon = 0$, for which 
  the instability occurs. The function $({\mng}/Kv)^2$
  is constant along the contours, and plotted in terms of the
  variables $(v_F/v)^2$ and $(h/2vm\gamma \ell)^2$. Outside the contours
  ${\mng}^2 < 0$.} 
  \label{fig}
\end{figure}

\begin{figure}
\begin{center}
  \includegraphics[width=.85\textwidth]{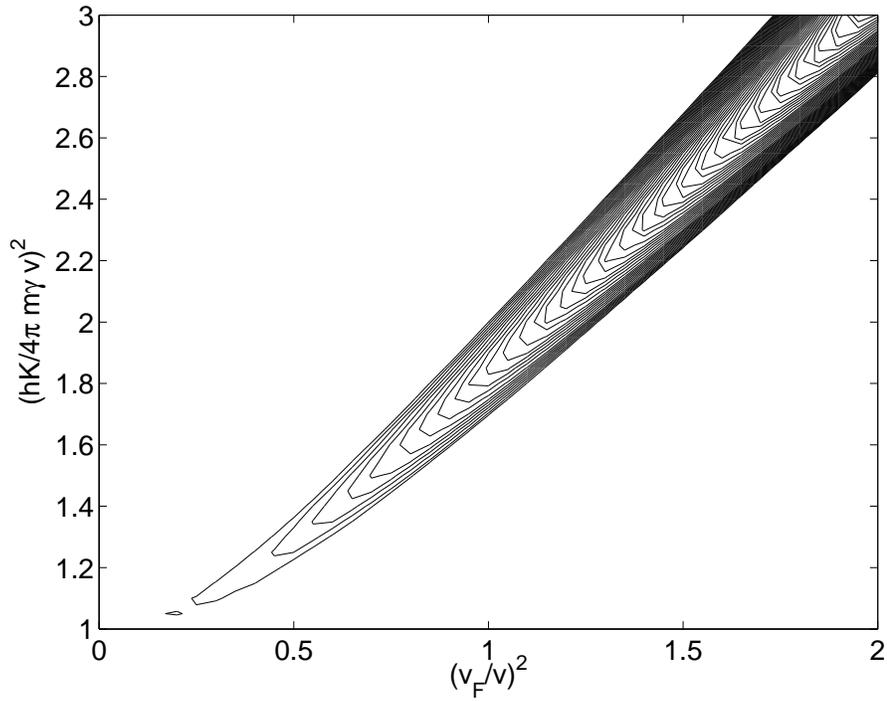}
\end{center}
  \caption{The same plot as in Fig.\ \ref{fig}, but for $\varepsilon =
  2$.}
  \label{fig2}
\end{figure}

Let us look at the simplest case of interacting neutrinos and
antineutrinos with  $M = N = 1$. We assume that they have equal
densities $n_0 = \bar{n}_0$, and are counter-propagating, 
i.e. $p_0 = -\bar{p}_0 > 0$. From (\ref{monopotential}) we obtain the
potential 
\begin{equation}
  V_{0\pm}(p) = \mp 4\sqrt{2}\,G_F\,{\sgn}(p)n_0
\end{equation}
while Eq.\ (\ref{monodispersion}) yields
\begin{multline}
  \shoveleft \sigma \left( \frac{\hbar K^2}{m\gamma} -2\sigma\varepsilon
  \right)\left\{ \left[ 
  \left({\mno} - \frac{p_0K}{m\gamma} \right)^2 - \left(
  \frac{\hbar K^2}{2m\gamma} \right)^2 -
  2\sigma\varepsilon\left(\left({\mno} - \frac{p_0K}{m\gamma}
  \right) - \frac{\hbar K^2}{2m\gamma}\right) \right]^{-1}
  \right. \\
  \shoveleft+ \left. \left[
  \left({\mno} + \frac{p_0K}{m\gamma} \right)^2 - \left(
  \frac{\hbar K^2}{2m\gamma} \right)^2 +
  2\sigma\varepsilon\left(\left({\mno} + \frac{p_0K}{m\gamma}
  \right) + \frac{\hbar K^2}{2m\gamma}
   \right) \right]^{-1} \right\} = 1,
\end{multline}
where $\sigma = 2\sqrt{2}\,G_F n_0/\hbar$ and $\varepsilon = 1 -
{\sgn}(p_0 - \hbar K) = 0$, $1$, or $2$ when $p_0 > \hbar K$, $p_0 =
\hbar K$, or $p_0 < \hbar K$, respectively. Thus, for the case
$\varepsilon = 0$, the growth rate is given by (see Figs.\ \ref{fig}
and  \ref{fig2})
\begin{equation}
  \frac{\mng^2}{K^2} = \sqrt{4v^2\left(\frac{\hbar
  K}{2m\gamma}\right)^2 + 4v^2v_F^2 + v_F^4} - v^2
  - v_F^2 - \left(\frac{\hbar K}{2m\gamma}\right)^2 ,
\label{monodispersion2}
\end{equation}
where ${\mng} = -i{\mno}$ is the instability
growth rate,  and $v_F^2 \equiv 2 \sqrt{2}\,G_F n_0/m\gamma $. 
Note also that, as expected, in the limit $v \rightarrow 0$, the
instability  disappears, just stating the well-known fact that there must be a
non-zero relative velocity between the beams in order for the instability to
occur.  As ${\mng}^2$ is positive, we have
\begin{equation}
  \left(\frac{v}{v_F}\right)^2 - 2 < \left(\frac{\hbar
  K}{2m\gamma v_F}\right)^2 < \left(\frac{v}{v_F}\right)^2
\label{stabcrit}
\end{equation}
i.e. 
\begin{equation}
  \ell_0 < \ell < \ell_0(1 - 2v_F^2/v^2)^{-1/2} ,  
\label{inequality}
\end{equation}
where we have introduced the length scales $\ell = 2\pi/K$ and $\ell_0
= \hbar/(2m\gamma v)$. 
Thus, a higher neutrino momentum can retain a smaller instability length
scale. It is clear from (\ref{monodispersion2}) that
(i) the instability will remain for arbitrary 
velocities (see Figs.~\ref{fig} and \ref{fig2}), and that (ii) the
higher the neutrino 
velocity, the smaller the corresponding instability length scale $\ell$. 

+

\subsection{Partial incoherence and thermal effects}

Partial incoherence can in general lead to lower growth rate, similar
to inverse Landau damping.  As an example of the results of stochastic
effects,  e.g.  thermal fluctuations, we look at the following example. 
Let the indeterminacy of the wave function manifest itself in a random
phase $\varphi(x)$ of the background wave function,
with the width $\Delta p$ defined according to $\langle
\mathrm{e}^{-i\varphi(x + y/2)} \mathrm{e}^{i\varphi(x - y/2)} \rangle
= \mathrm{e}^{-\Delta p|y|/\hbar}$. Due to this random spread, the
modulational instability will be damped, as will be shown below. The Wigner
function corresponding to the random phase assumption is given by the 
Lorentz distribution 
\begin{equation}
  f_0(p) = \frac{n_0}{\pi}\frac{\Delta p}{(p - p_0)^2 + \Delta p^2} . 
\end{equation}
With this, we obtain Eq.~(\ref{monodispersion2}) with
${\mng} \rightarrow {\mng}_D +
\Delta pK/m\gamma $, where ${\mng}_D$ is the reduced
growth rate. Thus, we see that the broadening tends to
suppress the growth. Moreover, a positive growth rate
${\mng}_D$ requires 
\begin{equation}
  \frac{2v\Delta p}{\hbar{\mng}} <
  \frac{\ell}{\ell_0} ,
  \label{damping}
\end{equation}
where ${\mng}$ is given by
Eq.~(\ref{monodispersion2}). Hence, the general property of a spread in
momentum space, here exemplified by a random phase, is to put bounds
on the modulational instability length scale $\ell$. 

\begin{figure}[t]
\begin{center}
  \includegraphics[width=\textwidth]{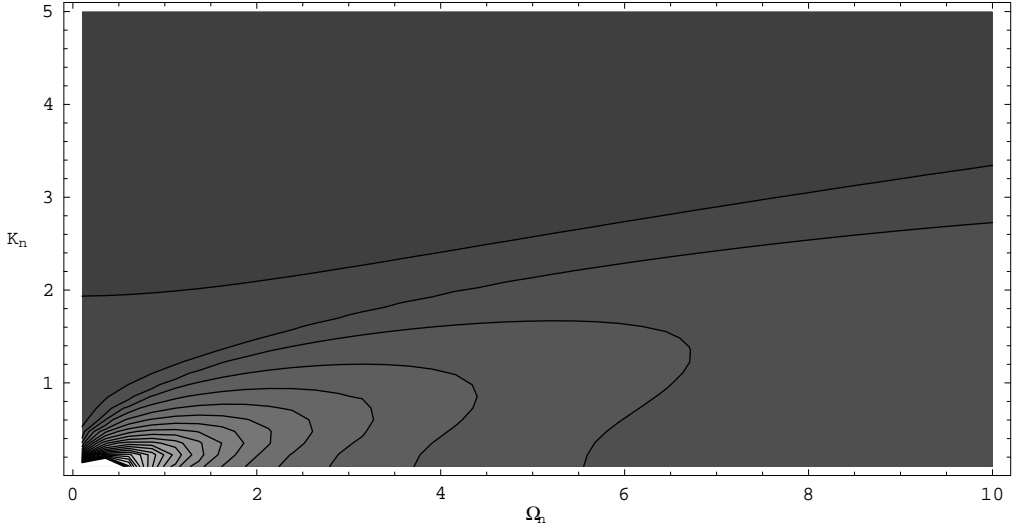}
\end{center}
  \caption{A contour plot of the Cauchy principal value
  $\mathrm{P}(I({\mno}_n, K_n))$ as a 
  function of the dimensionless variables ${\mno}_n$ and $K_n$. We
  note that $\mathrm{P}(I({\mno}_n, K_n)) \geq 0$, being largest for
  small ${\mno}_n$ and $K_n$, and approaching zero at infinity. The
  uppermost contour has $\mathrm{P}(I({\mno}_n, K_n)) = 0$.} 
  \label{fig3}
\end{figure}

\begin{figure}[t]
\begin{center}
  \includegraphics[width=\textwidth]{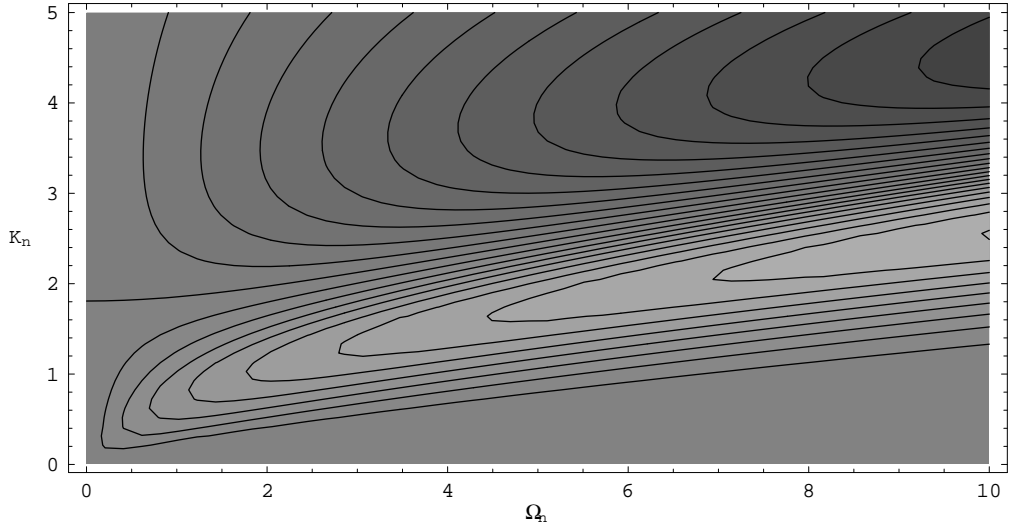}
\end{center}
  \caption{A contour plot of the contribution $g({\mno}_n, K_n)$ 
  of the poles to the integral (\ref{Fermi-Dirac-integral}) as a 
  function of the dimensionless variables ${\mno}_n$ and $K_n$. The
  darker areas represent negative values, the lighter positive
  values, and $g$ is zero on the contour emanating from $({\mno}_n, K_n) = (0,
  1.8)$. }
  \label{fig4}
\end{figure}

Incoherent effects among the neutrinos and anti-neutrinos can also 
be approached for a background obeying Fermi--Dirac statistics, i.e.
\begin{equation}
  f_{0\pm}(p) = \frac{c n_0}{\ln(4)k_B T_{\pm}}\left[ 1 + \exp\left(
  \frac{c|p|}{k_B T_{\pm}} \right)\right]^{-1},
\end{equation} 
where we set $M = N = 1$, and assume $n_0 = \bar{n}_0$. Here, we have 
neglected the mass of the neutrinos (which will give us
the correct result to lowest order). We will for simplicity assume
that $T_{\pm} = T$, so that the dispersion relation (\ref{dispersion2}) 
takes the form 
\begin{eqnarray}
  1 &=& -\frac{4\sqrt{2}\,c G_F n_0}{\ln(4)\hbar k_B
  T}\int_{-\infty}^{\infty}\,dp\, ({\mno} - pK/m\gamma)^{-1}\left[
  \left( 1 + \exp(c|p + \hbar K/2|/k_B T) \right)^{-1}
  \right. \nonumber \\  
  &&\qquad \qquad \qquad \left.
  + \left( 1 + \exp(c|p - \hbar K/2|/k_B T) \right)^{-1}  \right] .
\label{Fermi-Dirac-integral}
\end{eqnarray}
The dispersion relation (\ref{Fermi-Dirac-integral}) cannot be solved
analytically, but it can be expressed according to
\begin{eqnarray}
  1 &=& -Q \left[ \mathrm{P}(I({\mno}_n, K_n)) + i\pi g({\mno}_n,
  K_n) \right], 
\label{Fermi-Dirac-numerical}
\end{eqnarray}
where $\mathrm{P}(I({\mno}_n,K_n))$ is the principal value of
the integral 
\begin{equation}
  I = \int_0^{\infty} dx\, (1 + \e^x)^{-1} \left[ \frac{{\mno}_n +
  K_n^2}{({\mno}_n + K_n^2)^2 - K_n^2 x^2} + \frac{{\mno}_n -
  K_n^2}{({\mno}_n - K_n^2)^2 - K_n^2 x^2} \right] , 
\end{equation}
and $g = g_+ + g_-$, where
\begin{equation}
  g_{\pm}({\mno}_n, K_n) = 
  \frac{{\mno}_n \pm K_n^2}{1 + 
  \exp(|{\mno}_n \pm K_n^2|/\sqrt{2}K_n)}, 
\end{equation}
$Q \equiv (4/\ln 4)(2\sqrt{2}\,G_F n_0/k_B T)(m\gamma c^2/k_B T)$, and
we have introduced the dimensionless variables ${\mno}_n \equiv
(\hbar m\gamma c^2/(k_B T)^2){\mno}$ and $K_n \equiv (\hbar
c/\sqrt{2}\,k_B T) K$. The constant $Q$ gives the ratio of the
potential energy contribution of the background and the individual
neutrino energy to the thermal energy of the background. Furthermore,
$m\gamma c^2 \simeq k_B T$, thus simplifying the expression for $Q$. The
contributions from real and imaginary parts to the dispersion relation
are plotted in Figs.\ \ref{fig3} and \ref{fig4}, respectively. Note
that for very short length scales, i.e. large $K$, the quantity
${\mno}_n - K_n^2$ becomes negative, and the imaginary part in
Eq.\ (\ref{Fermi-Dirac-numerical}) change sign, something which will
not show using the long wavelength limit equation (\ref{Vlasov}). This
``quantum'' behaviour can in principle lead to growth instead of
damping of the perturbations (see Ref.\ \cite{Anderson2-etal} for a
general discussion of this behaviour). We can obtain a quantitative
measure of the growth rate as follows. For
any fixed $K_{n0}$, the dimensionless growth rate ${\mng}_n =
-i\,\mathrm{Im}\, {\mno}_n$ may 
be expressed as, denoting the value at ${\mno}_{n0}$ by $0$, 
\begin{equation}
  {\mng}_n = \pi
  \frac{\left( Q^{-1} + \mathrm{P}(I_0)
  \right)\,(\partial g/\partial{\mno}_{n0}) -
  g_0(\partial\mathrm{P}(I)/\partial{\mno}_{n0}) }{\pi^2\left(
  \partial g/\partial{\mno}_{n0} \right)^2 + \left(
  \partial\mathrm{P}(I)/\partial{\mno}_{n0} \right)^2 }
\label{Fermi-Dirac-growthrate}
\end{equation}
to first order around $({\mno}_{n0}, K_{n0})$. Thus, ${\mng}_n > 0$ if
$\left( Q^{-1} + \mathrm{P}(I_0) \right)\,(\partial \ln
g/\partial{\mno}_{n0}) >
\partial\mathrm{P}(I)/\partial{\mno}_{n0}$. Moreover, using values
given in Sec.\ \ref{sec:application}, one can show that $Q^{-1} \simeq
3\times10^9$. Thus, over a wide range of $({\mno}_n, K_n)$, $Q^{-1}$
dominates the contribution to the growth 
rate, and a positive growth rate is implied as long as
$\partial g/\partial{\mno}_{n0} > 0$.

\section{Applications}\label{sec:application}

As a model for hot dark matter, massive neutrinos have for some time
been one of the prime candidates, but as such they have faced the problem 
of the scale of the inhomogeneities they can support. Due to the conservation
of phase-space density, the Tremaine--Gunn limit constrains the
neutrino mass for isothermal spheres of a given size. For dwarf
galaxies, for which there are ample evidence of dark matter
\cite{Carr}, the necessary mass of the neutrino is uncomfortably large
\cite{Tremaine-Gunn,White-etal}. On the other hand, as pointed out by
Raffelt \& Silk \cite{Raffelt-Silk}, interacting dark matter can in
principle change this picture. Here we see  
from Eq.~(\ref{inequality}) that as the neutrino momentum increases, the
typical length scale $\ell$ of the inhomogeneity that can be
supported by the modulational instability \emph{decreases}. From the
definition of $\ell_0$, we note that as $v$ tends to $c$, $\ell_0
\rightarrow 0$, and due to Eq.~(\ref{inequality}) the allowed scale of
inhomogeneity becomes squeezed between two small values. On the other
hand, if $v \sim v_F$ (a condition stating that the neutrino number
density must reach extreme values), 
the upper inhomogeneity scale limit diverges.  
A minimum requirement for the effect to be of importance is
that the instability growth rate is larger than the Hubble
parameter $H$.  
An estimate of the growth rate can be obtained as follows. At the
onset of ``free streaming'' of neutrinos (i.e. their decoupling from
matter and radiation) at $z \sim 10^{10}$, 
the neutrino number density can be estimated as
$n_0 \simeq 2.1 \times 10^{38}\,\mathrm{m}^{-3}$ (see, e.g.
\cite{Peebles}). Furthermore, we assume that the  neutrino mass is in
the range $m \sim 1\,\mathrm{eV}$, and find $v_F \simeq
2.1\times10^6 \gamma^{-1/2}\,\mathrm{m}/\mathrm{s}$. The temperature
of the neutrinos, given by $T_{\nu} = (4/11)^{1/3}T_0(1 + z)$ ($T_0$
being the present day CMB temperature) \cite{Peebles}, at neutrino
decoupling is $T_{\nu} \simeq 2\times 10^{10}\,\mathrm{K}$. Thus, the
thermal energy is roughly five orders of magnitude greater than the
assumed rest mass of the neutrino, and in this sense the neutrinos can
be well approximated as ultra-relativistic.  
In this case, using values of $(\hbar K/2m\gamma  v_F)^2$ in the
middle range of the inequality (\ref{stabcrit}), Eq.\ 
(\ref{monodispersion2}) gives ${\mng} \simeq
2\sqrt{2}G_F n_0/\hbar \sim 16\times10^{10} \, \mathrm{s}^{-1}$ for the
values specified above. Assuming a critical density for the Universe,
the Hubble time becomes $H^{-1} \simeq H_0^{-1}(1+z)^{-3/2} \sim
5\times 10^2 \, \mathrm{s}$ at a redshift $10^{10}$, and thus
${\mng}/H \gg 1$. 

Although the two-stream instability may seem contrived as a
cosmological application, the important issue displayed by this
example is the non-gravitational growth of inhomogeneities, given a
small perturbation of a homogeneous, although anisotropic,
background. The fact that the growth rate exceeds the inverse of the
Hubble time by many orders of magnitude makes it clear that the
mechanism may be of some importance. Moreover, the analogous estimate
for the Fermi--Dirac background, although done in a simplistic manner,
indicates that the growth of the large $K$ perturbations may
be of importance. Note that this effect is a result of the use of the
\emph{full} Wigner--Moyal system, as compared to the Vlasov system
(\ref{Vlasov}), where these short wavelength effects are manifestly
neglected. Furthermore, it could also be of interest to use the current 
formalism as a tool to investigate neutrino interactions within supernov{\ae}, 
where the two-stream instability scenario may occur as a more natural
ingredient 
than perhaps within cosmology.

\section{Conclusion}

In conclusion, we have considered the nonlinear coupling between 
neutrinos and anti-neutrinos in a dense plasma. It is found that their
interactions are governed by a system of Wigner-Moyal equations,
which admit a modulational instability of the neutrino/antineutrino beams
against large scale (in comparison with the neutrino wavelength) 
density fluctuations. Physically, instability arises because interpenetrating 
neutrino and antineutrino beams are like quasiparticles, carrying free 
energy which can be coupled to inhomogeneities due to a resonant 
quasiparticle-wave interaction that is similar to a Cherenkov interaction.
Nonlinearly excited density fluctuations can be associated with the background 
inhomogeneity of the early Universe, and possibly counteract the free
streaming  smoothing of the small scale primordial fluctuations, thus making
massive neutrinos plausible as a candidate for hot dark matter.

\section*{Acknowledgements}

This research was partially supported by the Sida/NRF grant
SRP-2000-041 as well as by the Swedish Research Council 
through Contract No. 621-2001-2274 and the Deutsche 
Forschungsgemeinschaft through the Sonderforschungsbereich 591.

\end{document}